\documentclass[a4paper,fleqn,usenatbib]{mnras}

\usepackage{newtxtext,newtxmath}
\usepackage[T1]{fontenc}
\usepackage{ae,aecompl}

\usepackage{graphicx}	% Including figure files
\usepackage{amsmath}	% Advanced maths commands
\usepackage{amssymb}	% Extra maths symbols

\title[IGM baryon mass fraction from FRB observations]{Cosmology-insensitive estimate of IGM baryon mass fraction from five localized fast radio bursts}

\author[Li et al.]{Z. Li$^1$, 
	H. Gao$^1$\thanks{E-mail: gaohe@bnu.edu.cn}
	J.-J. Wei$^2$, 
	Y.-P. Yang$^3$,
	B. Zhang$^4$,
	and Z.-H. Zhu$^1$\thanks{E-mail: zhuzh@bnu.edu.cn}
	\\
	$^{1}$Department of Astronomy, Beijing Normal University, Beijing 100875, China \\
	$^{2}$Purple Mountain Observatory, Chinese Academy of Sciences, Nanjing 210034, China\\
	$^{3}$South-Western Institute for Astronomy Research, Yunnan University, Kunming 650500, China\\
	$^{4}$Department of Physics and Astronomy, University of Nevada, Las Vegas, NV 89154, USA\\}

\begin{document}
	\label{firstpage}
	\pagerange{\pageref{firstpage}--\pageref{lastpage}}
	\maketitle
	
	\begin{abstract}
		Five fast radio bursts (FRBs), including three apparently non-repeating ones FRB 180924, FRB 181112, and FRB 190523, and two repeaters, FRB 121102 and FRB 180916.J0158+65, have already been localized so far. We apply a method developed recently by us (Li et al. 2019) to these five localized FRBs to give a cosmology-insensitive estimate of the fraction of baryon mass in the IGM, $f_{\rm IGM}$. Using the measured dispersion measure (DM) and luminosity distance $d_{\rm L}$ data (inferred from the FRB redshifts and $d_{\rm L}$ of type Ia supernovae at the same redshifts) of the five FRBs, we constrain the local $f_{\rm IGM} = 0.84^{+0.16}_{-0.22}$ with no evidence of redshift dependence. This cosmology-insensitive estimate of $f_{\rm IGM}$ from FRB observations is in excellent agreement with previous constraints using other probes. Moreover, using the three apparently non-repeating FRBs only we get a little looser but consistent result $f_{\rm IGM} = 0.74^{+0.24}_{-0.18}$. In these two cases, reasonable estimations for the host galaxy DM contribution (${\rm DM_{host}}$) can be achieved by modelling it as a function of star formation rate. The constraints on both $f_{\rm IGM}$ and ${\rm DM_{host}}$ are expected to be significantly improved with the rapid progress in localizing FRBs.
	\end{abstract}
	
	
	%%%%%%%%%%%%%%%%%%%%%%%%%%%%%%%%%%%%%%%%%%%%%%%%%%
	
	%%%%%%%%%%%%%%%%% BODY OF PAPER %%%%%%%%%%%%%%%%%%
	
	\section{Introduction}
	Fast radio bursts (FRBs) are intense bursts of radio waves with millisecond durations \citep{lorimer07,thornton13,petroff15,petroff16}. Their large values of the dispersion measure (DM),  which are much greater than expected contributions from the Milky Way interstellar medium alone \citep{dolag15}, suggest their extragalactic or even cosmological origin. Therefore, the observed DM of an FRB should at least have a significant contribution from the intergalactic medium (IGM). It is implied that, based on the ${\rm DM_{IGM}}-z$ relation, the FRBs are typically at distances of the order of
	gigaparsec \citep{ioka03,inoue04,deng14,zheng14,zhang18}. The cosmological origin of FRBs allows them to become promising tools for studying the universe and fundamental physics, e.g. baryon density and spatial distribution \citep{mcquinn14,deng14,munoz18}, dark energy \citep{gao14, zhou14, walters18,wei18}, cosmic ionization history \citep{deng14,zheng14}, the large-scale structure of the universe \citep{masui15}, the Einstein's equivalence principle \citep{wei15, nusser16, tingay16}, the rest mass of the photon \citep{wu16,shao17}, the cosmic proper distance measurements \citep{wang17}, as well as constraints on the magnetic fields in the IGM \citep{akahori16}. Because of another two properties of these mysterious transients, i.e. short durations ($\sim(1-10) \mathrm{ms}$) \citep{lorimer07, keane11, thornton13, spitler14, petroff15, ravi15, champion16, petroff16, spitler16} and high event rate ($\sim10^3-10^4$ per day all sky) \citep{thornton13, champion16}, lensed FRBs have been proposed as a probe of compact dark matter \citep{munoz16, wang18}, motion of the FRB source \citep{dai17}, and precision cosmology \citep{li18,liu19}.
	
	More recently, \citet{li19} proposed a method to probe the intergalactic medium (IGM) by estimating the fraction of baryon in the IGM, $f_{\rm IGM}$, using the putative sample of FRBs with both of their DMs and luminosity distances $d_{\rm L}$ available. It should be emphasized that, since $d_{\rm L}/{\rm DM}$ essentially does not depend on cosmological parameters, this method can determine $f_{\rm IGM}$ in a cosmology-insensitive way. By means of Monte Carlo simulations, it was shown that an unbiased and cosmology-insensitive estimate of the present value of $f_{\rm IGM}$ can be obtained from a sample of FRBs with joint measurements of DM and $d_{\rm L}$. In addition, such a method can also simultaneously lead to a measurement of the mean value of DM contributed from the local host galaxy. In practice, it should be also pointed out that localization and redshift measurement of bursts play an essential role for applying this approach. Back then, localization was only made for the first repeater FRB 121102, which has $z \sim 0.19$ \citep{spitler16,scholz16,chatterjee17,marcote17,tendulkar17,zhang19,wang19}. 
	
	Recently, another three apparently non-repeating FRBs\footnote{Since all FRBs can be in principle repeating, here we use "apparently non-repeating FRBs" to refer to the sources from which only one burst was detected so far. In the following text, This population is also referred as non-repeating or single bursts.} and one repeating source were localized. First, FRB 180924 was localized with the Australia Square Kilometer Array Pathfinder (ASKAP) to a position 4 kpc from the center of a luminous and massive galaxy at redshift 0.3214 \citep{bannister19}. The observed DM of the burst, is $361.42\pm0.06~{\rm pc~cm^{-3}}$ and the pulse width is $1.30\pm0.09~{\rm ms}$. The position of the burst is in one of the high Galactic latitude fields targeted by ASKAP (Galactic longitude $b=0.742467^{\circ}$ , Galactic latitude $l=-49.414787^{\circ}$ ). The Milky Way DM disk and halo components estimated from the NE2001 model \citep{cordes02} for this position are $40.5~{\rm pc~cm^{-3}}$ and $30~{\rm pc~cm^{-3}}$, respectively. Second, FRB 190523 was localized by the Deep Synoptic Array ten-antenna prototype (DSA-10) to a few-arcsecond region containing a single massive galaxy at redshift 0.66 \citep{ravi19}. The burst was detected at a DM of $760.8\pm0.6~{\rm pc~cm^{-3}}$ and the pulse width is $0.42\pm0.05~{\rm ms}$. The position of the burst is Galactic longitude $b=117.03^{\circ}$, Galactic latitude $l=44^{\circ}$. In this direction, the Milky Way disk and the Milky Way ionised halo contribute $37~{\rm pc~cm^{-3}}$ and $50-80~{\rm pc~cm^{-3}}$, respectively. Therefore, the extragalactic DM of this burst is between 644 and $674~{\rm pc~cm^{-3}}$. Third, FRB 181112 was detected by the Commensal Real-time ASKAP Fast Transients (CRAFT) survey~\citep{prochaska19b}. The burst sweep yields an estimate of the FRB dispersion measure ${\rm DM} = 589.27\pm0.03~{\rm pc~cm^{-3}}$. The real-time detection precisely localized the burst to a sky position Galactic longitude $b=342.6^{\circ}$, Galactic latitude $l=-47.7^{\circ}$. The Milky Way disk and the Milky Way ionised halo contribute $102~{\rm pc~cm^{-3}}$ and $\sim 30~{\rm pc~cm^{-3}}$ in this sky position. The image centered on FRB 181112 obtained with the FOcal Reducer/low dispersion Spectrograph 2 (FORS2) instrument on the Very Large Telescope (VLT) shows the presence of a galaxy coincident with this burst, previously cataloged by the Dark Energy Survey (DES) as DES J214923.66-525815.28. More interestingly, the DES and FORS2 data also show that a luminous galaxy (DES J214923.89-525810.43) is $\sim5''$ to the North of the FRB event. Follow-up spectroscopy with the FORS2 instrument obtained that the redshift of the former galaxy (host galaxy of this burst) is $z = 0.47550$, and the redshift of the latter (i.e. the foreground galaxy) is $z = 0.3674$. This foreground galaxy contributes $50-120~{\rm pc~cm^{-3}}$, which does not dominate the DM of this event and depends on assumptions of density profile and the total mass of the halo gas. In our following analysis, we approximately take $85\pm35~{\rm pc~cm^{-3}}$ to account for the DM contribution and corresponding uncertainty of this foreground galaxy. Finally, a second repeater, FRB180916.J0158+65, was localized recently to be located in a star-forming region of a nearby massive galaxy at $z=0.0337$~\citep{marcote20}. In the direction of this burst, the Milky Way disk (halo) contributes $199~(50-80)~{\rm pc~cm^{-3}}$.

	In the following, we will apply the method of \cite{li19} to perform a cosmological-model-insensitive estimate of the fraction of baryons in the IGM using the five FRBs with redshift measurements.

	\section{Methods}
	The method has been proposed in \cite{li19} and below we briefly summarize it.
	
\begin{figure*}
		\centering
		\includegraphics[width=0.45\textwidth]{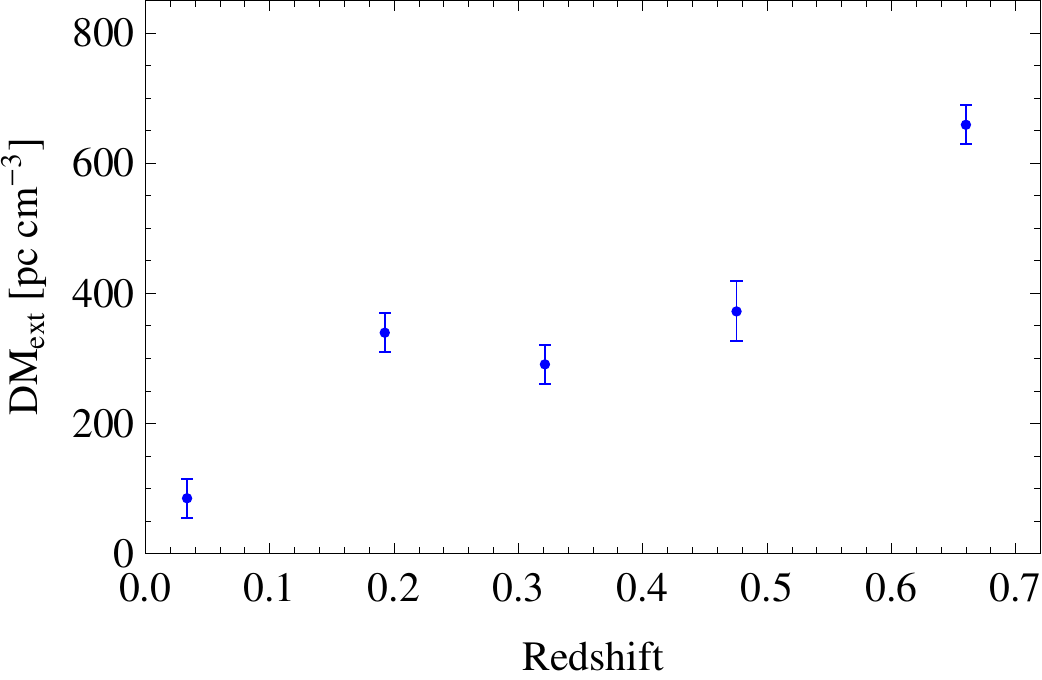}
		\includegraphics[width=0.45\textwidth]{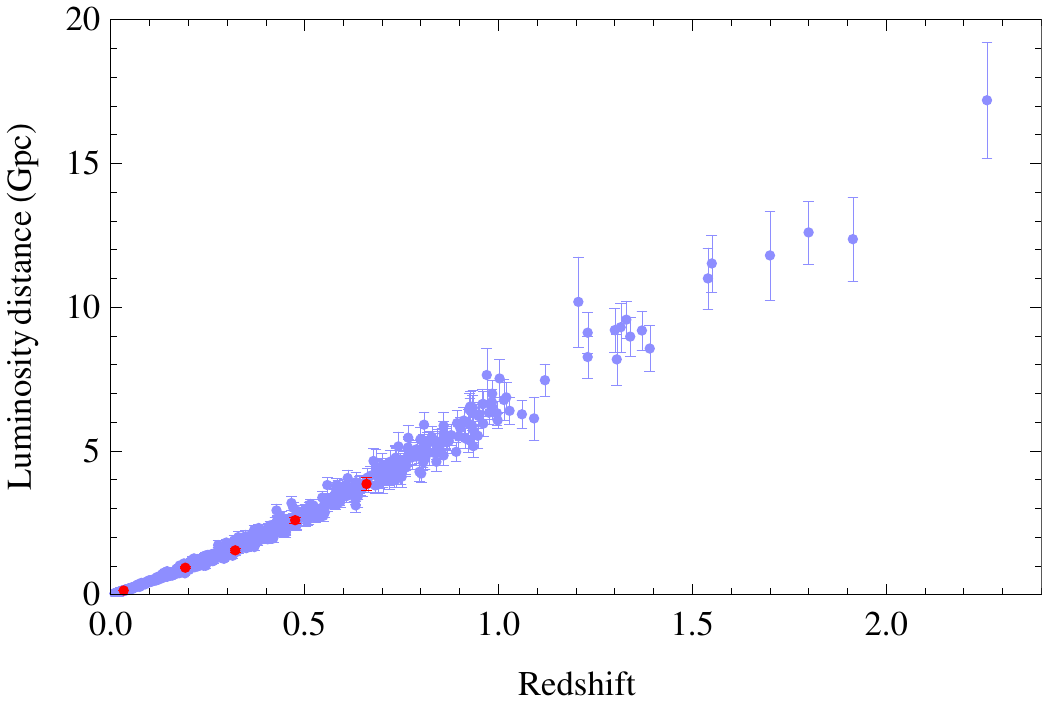}
		\caption{\label{Fig1}{\bf Left:} Extragalactic dispersion measures of all five localized FRBs. {\bf Right:} The Hubble diagram of Pantheon SNe Ia and luminosity distances binned from the two nearest SNe Ia for five FRBs with their host galaxy identified.}
	\end{figure*}

	For a localized FRB, we can get the observed extragalactic DM with a considerable precision by deducting Milky Way portion, $\rm{DM_{MW}}$, from $\rm{DM_{obs}}$ \citep{thornton13,deng14,yang16}.
	\begin{equation}\label{eq1}
	\rm{DM_{ext,obs}}=\rm{DM_{obs}}-\rm{DM_{MW}}.
	\end{equation}
	Its corresponding uncertainty is $\sigma_{\rm{ext}}=\sqrt{\sigma_{\rm{obs}}^2+\sigma^2_{\rm{MW}}}$. For sources at high Galactic latitude ($|b| >10^{\circ}$), the average uncertainty of the DM contribution from the Milky Way and halo, $\sigma_{\rm{MW}}$, is about $30~{\rm {pc~cm^{-3}}}$ \citep{manchester05}. $\rm{DM_{ext,obs}}$ and $\sigma_{\rm{ext}}$ for all five localized FRBs are shown in the left panel of Fig. \ref{Fig1}. Moreover, the observed extragalactic DM actually consists of two contributions, i.e. the IGM portion and the host galaxy portion. Therefore, for a localized FRB, $\rm{DM}_{\rm{IGM}}$ in the certain direction can be in principle measured by deducting the host galaxy portion from the observed extragalactic DM \citep{ioka03,deng14}
	\begin{equation}\label{eq2}
	{\rm DM_{IGM,obs}}={\rm DM_{ext,obs}}-\frac{{\rm DM_{host,loc}}}{1+z}.
	\end{equation}
	The value of $\rm {DM_{host,loc}}$ and its uncertainty are intractable parameters since they are poorly known and dependent on many factors, e.g. the type of the host galaxy, the inclination angle of the galaxy disk, the site of FRB in the host, and the near-source plasma \citep{luan14,katz16,piro16,yang16,metzger17,piro17,yangzhang17,yang17}. For a given type of host galaxy, the magnitude of $\rm {DM_{host,loc}}$ along a certain line of sight depends on the electron density and size of the galaxy \citep{manchester06,xu15}. The former one is proportional to the H$\alpha$ luminosity of the galaxy, which scales with the star formation rate (SFR) \citep{kennicutt94,madau98}. Therefore, for simplicity and generality, here we model $\rm {DM_{host,loc}}$ as a function with respect to redshift by assuming that the rest-frame DM distribution accommodates the evolution of the SFR history \citep{luo18}, ${\rm DM_{host, loc}}(z)={\rm DM_{host, loc, 0}}\sqrt{{\rm SFR}(z)/{\rm {SFR}(0)}}$, and adopt a continuous form of a broken power law for it \citep{hansan08}. In our following analysis, ${\rm DM_{host, loc, 0}}$ is treated as a free parameter to be fitted. In this case, we write the uncertainty of ${\rm DM_{IGM}}$ as 
	\begin{equation}\label{eq3}
	\sigma^2_{\rm{IGM}}=\sigma_{\rm{obs}}^2+\sigma^2_{\rm{MW}}+\sigma^2_{\rm int}.
	\end{equation}
	For the sample of these five localized FRBs, as shown in Fig. \ref{Fig1}, observed ${\rm DM_{ext}}$ has significant scatter. This systematical scatter might originate from the diversity of host galaxy contribution or the IGM fluctuation. Therefore, we introduce $\sigma_{\rm int}$ in Eq. \ref{eq3} to account for this scatter and the magnitude of it is determined in the following $\chi^2$-statistics analysis by ensuring the reduced $\chi^2$ ($\chi^2/dof$, $dof$: degree of freedom) is approximately to be unity.
	
	In theory, the IGM portion depends on the cosmological distance scale the burst propagates through and the fraction of ionized electrons in hydrogen and helium on the path. If both hydrogen and helium are fully ionized (valid below $z\sim3$), it can be expressed as
	\begin{equation}\label{eq4}
	\mathrm{DM}_{\mathrm{IGM}}(z)=\frac{21cH_0\Omega_b}{64\pi G m_{\mathrm{p}}}\int_0^z\frac{f_{\mathrm{IGM}}(z')(1+z')dz'}{\sqrt{\Omega_m(1+z)^3+\Omega_\Lambda}}.
	\end{equation} 
	It is known that $f_{\rm{IGM}}$ grows with redshift as massive haloes are more abundant in the late universe \citep{mcquinn14, prochaska19a}. Therefore, we characterize the growth of $f_{\rm{IGM}}$ with redshift as a mildly increasing function, $f_{\rm{IGM}}(z)=f_{\rm{IGM,0}}(1+\alpha z/(1+z))$ \citep{li19}. Here, the parameter $\alpha$ is responsible for quantifying the evolution of $f_{\rm{IGM}}$. The fact that $f_{\rm{IGM}}$ grows with redshift requires $\alpha>0$.
	
	The ${\rm DM_{IGM}}-z$ relation has been widely used for probing baryon density and cosmography \citep{mcquinn14,deng14,gao14, zhou14, walters18,wei18}. However, almost all these works have encountered the degeneracy between the fraction of baryon in the IGM, $f_{\rm IGM}$, and cosmological parameters. This degeneracy is intractable when the standard cosmological model itself is very uncertain, which is strongly implied by the well-known Hubble constant tension \citep{planck18,riess19}. Therefore, in order to alleviate the dependence of estimation for $f_{\rm IGM}$ from the ${\rm DM_{IGM}}-z$ relation on cosmology, the combination of DM and luminosity distance measurements of FRBs, $d_{\rm L}/{\rm DM_{IGM}}$, has been proposed to estimate $f_{\rm IGM}$ in a cosmological-model-insensitive way \citep{li19}\footnote{In \citep{li19}, we have concluded that, compared to the ${\rm DM_{IGM}}-z$ relation, $d_{\rm L}/{\rm DM_{IGM}}$ is insensitive to the dark energy equation of state parameter, $w$, by a factor of 10. Here, we further check the sensitivities of the ${\rm DM_{IGM}}-z$ relation and $d_{\rm L}/{\rm DM_{IGM}}$ to cosmology by considering variations in the matter density parameter, $\Omega_m$, and the same conclusion is obtained. It should be pointed out that, in this work, the term ``cosmology-insensitive" only implies that results here are insensitive to cosmological-model parameters, such as $\Omega_m$ and $w$.} By denoting $d_{\rm L}/{\rm DM_{IGM}}=R$, we obtain estimates for parameters by carrying out the following $\chi^2$ analysis
	\begin{equation}\label{eq5}
	\chi^2 = \sum_i \frac{ ( R_{{\rm obs},i} - R_{{\rm th},i} )^2 }{ \sigma_{{\rm tot},i}^2},
	\end{equation}
	where $R_{\rm obs} = d_{\rm L,obs} /{\rm DM_{IGM, obs}}$, $R_{\rm th} = d_{\rm L,th} / {\rm DM_{IGM,th}}$, and 
	\begin{equation}\label{eq6}
	\sigma^2_{\rm tot} = \frac{\sigma^2_{d_{\rm L,obs}}}{\rm DM^2_{IGM,obs}}+\frac{d^2_{\rm L,obs}\cdot\sigma^2_{\rm IGM}}{\rm DM^4_{IGM,obs}}.
	\end{equation}
	
	Here, for each localized FRB, we obtain its distance by binning the observed distances of the two SNe Ia with redshifts closest to the five FRBs. The Hubble diagram of the Pantheon SN Ia \citep{scolnic18} and binned distances for the latest localized FRBs are shown in the right panel of Fig. \ref{Fig1}. Using the method proposed in \citep{li19}, we apply the observational data of the five localized FRBs to perform a cosmology-insensitive estimate of $f_{\rm IGM}$.

	\section{Results}
\begin{figure}
		\centering
		\includegraphics[width=0.48\textwidth]{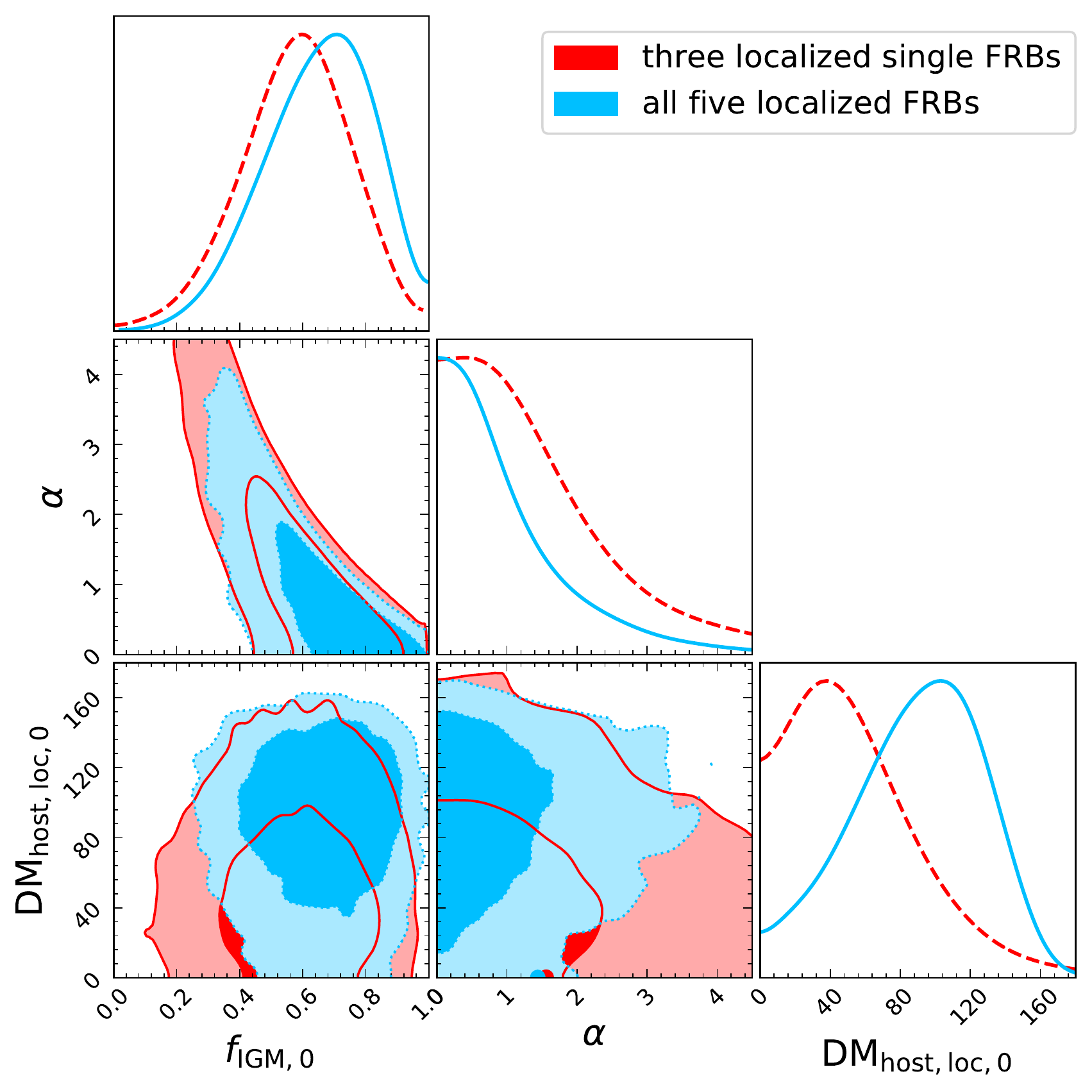}
		\caption{Joint contours for all fitted parameters from the latest localized FRBs. }
		\label{Fig2}
	\end{figure}

	We first obtain the constraints from all five FRBs with redshift measurements. The results are shown in Fig. \ref{Fig2} (light blue contours).  In this case, we obtain that the present value of $f_{\rm IGM}$ is constrained to be $f_{\rm IGM,0}=0.84^{+0.16}_{-0.22}$. This is well consistent with previous results obtained from observations \citep[e.g.][]{fukugita98,fukugita04,shull12,hill16,munoz18} or simulations \citep[e.g.][]{cen99,cen06}, $f_{\rm IGM}\sim0.83$). In addition, with the requirement that $f_{\rm IGM}$ grows with redshift ($\alpha>0$), $\alpha$ is estimated to be $\alpha<4.0$ at 2$\sigma$ confidence level, which implies that there is no strong evidence for the redshift dependence of $f_{\rm IGM}$. Meanwhile, an estimate for the value of DM contributing from host galaxies, ${\rm DM_{host,loc,0}} =107^{+24}_{-45}~{\rm pc~cm^{-3}}$, is achieved. This is likely due to the inclusion of FRB 121102 whose ${\rm DM_{host,loc,0}}$ is known to be very large, i.e. 55 to 225 ${\rm pc~cm^{-3}}$ \citep{tendulkar17,bassa17}. In this $\chi^2$ fitting analysis, we also obtain an estimation for the average IGM fluctuation in directions of these five FRBs, $\sigma_{\rm int}\simeq40~{\rm pc~cm^{-3}}$. That is, for these five localized FRBs, the systematical scatter due to IGM fluctuation is much smaller than that predicted by simulations in \citet{mcquinn14,dolag15}. Secondly, since the population of single bursts might be different from the one of repeaters\footnote{Note that whether or not all FRBs have repeating nature is still under debate \citep{palaniswamy18,caleb19,james19}. \citet{ravi19na} showed that the volumetric occurrence rate of FRBs that have not been observed to repeat so far is larger than the rates of candidate cataclysmic progenitor events, suggesting that most observed FRBs may originate from repeating sources.}, we also constrain $f_{\rm IGM}$ by considering the three localized single bursts. Results are also shown in Fig. \ref{Fig2} (red contours). It  suggests that a smaller $f_{\rm IGM,0}=0.74^{+0.24}_{-0.18}$ and a larger $\alpha=0.89^{+1.41}_{-0.89}$ are obtained. These results are also in good consistency with what obtained in \citep{fukugita98,fukugita04}. Moreover, constraint on $\alpha$ slightly suggests the redshift dependence of $f_{\rm IGM}$. At the same time, an estimation for the value of host galaxy contribution DM is ${\rm DM_{host,loc,0}} =34^{+39}_{-31}~{\rm pc~cm^{-3}}$, which is in good agreement with what inferred from observations of the latest localized single bursts \citep{bannister19,ravi19}. It should be clarified that estimations for the host galaxy contribution DM are based on the assumption for the redshift dependence of the rest-frame DM distribution on the evolution of the SFR history.
	
\begin{figure}
		\centering
		\includegraphics[width=0.45\textwidth]{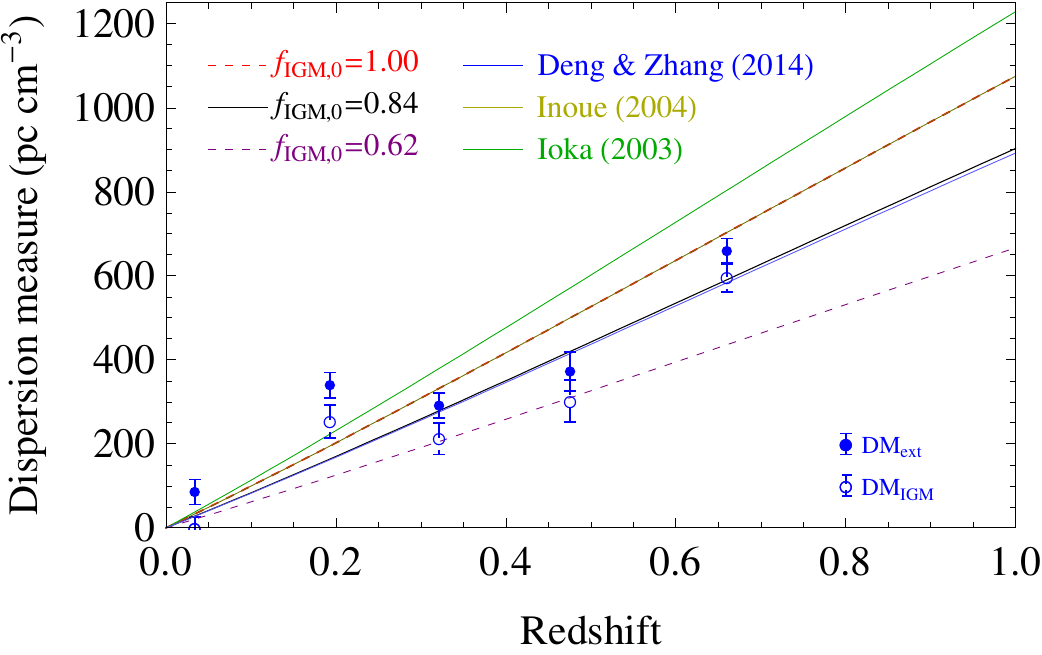}
		\caption{Extragalactic DM and IGM DM in contrast with ${\rm DM_{IGM}}-z$ predictions from different $f_{\rm IGM,0}$.}
		\label{Fig3}
	\end{figure}

	Taking the constraints from all five localized FRBs into consideration, in Fig. \ref{Fig3}, we plot extragalactic DM and IGM DM in contrast with ${\rm DM_{IGM}}-z$ predictions from different $f_{\rm IGM,0}$ in the $1\sigma$ range. It is found that derived ${\rm DM_{ext}}$ of all five bursts are above the best-fit (black solid) line. After deducting ${\rm DM_{host}}/(1+z)$ from ${\rm DM_{ext}}$, ${\rm DM_{IGM}}$ of all five bursts are evenly distributed around the best-fit line. These results suggest that the estimated $f_{\rm IGM}$ and ${\rm DM_{host,loc,0}}$ are reliable. In addition, some ${\rm DM_{IGM}}-z$ predictions, which have been widely used in the literature, are also plotted for the sake of comparison. It is obtained that the ${\rm DM_{IGM}}-z$ relation given by \citep{deng14}, ${\rm DM} \sim 855z~{\rm pc~cm^{-3}}$ \citep{zhang18}, with the consideration of He ionization history and $f_{\rm IGM}=0.83$ is in excellent agreement with cosmology-insensitive estimations from the latest observations. However, the relation with $f_{\rm IGM}=1.0$ given in \citep{inoue04}, ${\rm DM} \sim 1000z~{\rm pc~cm^{-3}}$, is marginally outside the $1\sigma$ range of the constraint. Furthermore, the relation with $f_{\rm IGM}=1.0$ and without considering He ionization history given in \citep{ioka03}, ${\rm DM} \sim 1200z~{\rm pc~cm^{-3}}$, is in significant tension with the latest FRB observations.

	\section{Conclusions and discussions}
	In this paper, we investigate cosmology-insensitive constraints on the fraction of baryons in the IGM, $f_{\rm IGM}$, from joint measurements of luminosity distances and dispersion measures of the latest localized FRBs, including the two repeaters, i.e. FRB 121102 and 180916.J0158+65, and three single bursts, i.e. FRB 180924, 181112, and FRB 190523. We estimate $f_{\rm IGM}$ using observations of all these five bursts and obtain $f_{\rm IGM,0}=0.84^{+0.16}_{-0.22}$. This constraint from FRB observations is well consistent with those obtained from other probes \citep{fukugita98,fukugita04,shull12,hill16}. {For the three localized single FRBs only, we obtain $f_{\rm IGM,0}=0.74^{+0.24}_{-0.18}$, which is looser but still consistent with the five-FRB constraint as well as the constraints derived from previous methods. For these cosmology-insensitive estimations, it is also necessary to point out that they are obtained with the best-fit cosmology from the latest $Planck$ CMB observations. In fact, parameters in this underlying cosmology also degenerate with those we are interested here and the complete set of parameters should be $\{H_0,\Omega_b,f_{\rm IGM,0},\alpha, {\rm DM_{host,loc,0}}\}$. However, these parameters would be very loosely constrained from only five localized FRBs. Therefore, at this moment, we focus on the estimate of IGM baryon mass fraction, which is one the most promising benefits of localized FRBs, by taking $H_0$ and $\Omega_b$ from the latest $Planck$ CMB observations as priors. Alternatively, all concerning parameters would be well constrained from a large number of localized FRBs which are likely to be achieved in the near future.
		
		Redshift measurements or localizations of a large sample of FRBs are essential for using this method to perform cosmology-insensitive estimate of $f_{\rm IGM}$. A substantial population of repeating FRBs will be detected by wide-field sensitive radio telescopes, such as the CHIME/FRB instruments \citep{chime19a,chime19b,chime19c}. It suggests that a considerable number of repeaters with redshifts measured will be accumulated in the very near future. It is more exciting that radio interferometers, such as ASKAP and DSA-10, has led to direct localizations of three single bursts. Based on the detection rate, it is likely to collect a large number of localized FRBs in a short time. As suggested in \citep{hallinan19}, a world-leading radio survey telescope and multi-messenger discovery engine for the next decade, DSA-2000, will simultaneously detect and localize $\sim 10^4$ fast radio bursts each year. It is foreseen that the constraints on concerning parameters, especially on $f_{\rm IGM}$, will be significantly improved with the rapid progress in localizing FRBs.

		\section*{Acknowledgements}
		
		We dedicate this article to the 60th anniversary of the Department of Astronomy of Beijing Normal University, the second astronomy programme in the modern history of China. We are grateful to Ye Li for her helpful discussions. We also would like to thank the referee for his/her constructive comments that have allowed us to improve the manuscript significantly. This work was supported by the National Natural Science Foundation of China under Grants Nos. 11722324, 11603003, 11633001, U1831122, and 11920101003, the Strategic Priority Research Program of the Chinese Academy of Sciences, Grant No. XDB23040100, and the Interdiscipline Research Funds of Beijing Normal University.
		
		%%%%%%%%%%%%%%%%%%%%%%%%%%%%%%%%%%%%%%%%%%%%%%%%%%
		
		
		
		%\bibliographystyle{mnras}


		%%%%%%%%%%%%%%%%%%%%%%%%%%%%%%%%%%%%%%%%%%%%%%%%%%
		
		%%%%%%%%%%%%%%%%% APPENDICES %%%%%%%%%%%%%%%%%%%%%
		
		%\appendix
		
		
		
		%%%%%%%%%%%%%%%%%%%%%%%%%%%%%%%%%%%%%%%%%%%%%%%%%%

		% Don't change these lines
		\bsp	% typesetting comment
		\label{lastpage}
	\end{document}